\documentclass[]{pasj02} 

\usepackage{threeparttable} 

\usepackage{comment}

\jyear{2024}
\Received{}
\Accepted{}

\begin{document} 

\title{ Time evolution of white-light flare accompanied by probable postflare loops on M-type dwarf EV Lacertae}

\author{
 Shinnosuke \textsc{Ichihara},\altaffilmark{1}\altemailmark\orcid{0009-0004-5808-4662} \email{ichihara@kusastro.kyoto-u.ac.jp} 
 Daisaku \textsc{Nogami},\altaffilmark{1}\orcid{0000-0001-9588-1872}
 Kosuke \textsc{Namekata},\altaffilmark{2,3,4,5,6}\orcid{0000-0002-1297-9485}
 Hiroyuki \textsc{Maehara},\altaffilmark{7}\orcid{0000-0003-0332-0811}
 Yuta \textsc{Notsu},\altaffilmark{8,9}\orcid{0000-0002-0412-0849}
 Kai \textsc{Ikuta},\altaffilmark{10}\orcid{0000-0002-5978-057X}
 Satoshi \textsc{Honda},\altaffilmark{11}\orcid{0000-0001-6653-8741}
 Takato \textsc{Otsu},\altaffilmark{12}\orcid{0009-0004-2275-3991}
 and 
 Kazunari \textsc{Shibata}\altaffilmark{13,14}\orcid{0000-0003-1206-7889}
}
\altaffiltext{1}{Department of Astronomy, KyotoUniversity, Kitashirakawa-Oiwake-cho, Sakyo-ku, Kyoto, Kyoto 606-8502, Japan}
\altaffiltext{2}{The Hakubi Center for Advanced Research, Kyoto University, Yoshida-Honmachi, Sakyo-ku, Kyoto 606-8501, Japan}
\altaffiltext{3}{Heliophysics Science Division, NASA Goddard Space Flight Center, 8800 Greenbelt Road, Greenbelt, MD 20771, USA}
\altaffiltext{4}{The Catholic University of America, 620 Michigan Avenue, N.E. Washington, DC 20064, USA}
\altaffiltext{5}{Division of Science, National Astronomical Observatory of Japan, NINS, Osawa, Mitaka, Tokyo, 181-8588, Japan}
\altaffiltext{6}{Department of Physics, Kyoto University, Kitashirakawa-Oiwake-cho, Sakyo-ku, Kyoto, 606-8502, Japan}
\altaffiltext{7}{Okayama Branch Office, Subaru Telescope, National Astronomical Observatory of Japan, NINS, Kamogata, Asakuchi, Okayama 719-0232, Japan}
\altaffiltext{8}{Laboratory for Atmospheric and Space Physics, University of Colorado Boulder, 3665 Discovery Drive, Boulder, CO 80303, USA}
\altaffiltext{9}{National Solar Observatory, 3665 Discovery Drive, Boulder, CO 80303, USA}
\altaffiltext{10}{Department of Social Data Science, Hitotsubashi University, 2-1 Naka, Kunitachi, Tokyo 186-8601, Japan}
\altaffiltext{11}{Nishi-Harima Astronomical Observatory, Center for Astronomy, University of Hyogo, 407-2, Nishigaichi,
 Sayo-cho, Sayo, Hyogo 679-5313, Japan}
\altaffiltext{12}{ Astronomical Observatory, Kyoto University, Sakyo-ku, Kyoto, Japan}
\altaffiltext{13}{ Kwasan Observatory, Kyoto University, Yamashina, Kyoto 607-8471, Japan}
\altaffiltext{14}{ School of Science and Engineering, Doshisha University, Kyotanabe, Kyoto 610-0321, Japan}



\KeyWords{stars: activity --- stars: flare --- stars: late-type}  

\maketitle

\begin{abstract}
 White-light flares are explosive phenomena accompanied by brightening of continuum from near-ultraviolet to optical, which occur on the Sun and stars. In order to investigate the mechanism of white-light flares, we carried out simultaneous optical photometry (TESS : 6000-10000 \AA) and spectroscopy (Seimei Telescope : 4100-8900 \AA) of the active flaring M-dwarf EV Lacertae on 2019 September 14. We detected a flare with high-time-cadence ($\sim \ 50$ sec) spectroscopic observation. At the peak, the continuum of the flare component is well fitted by a blackbody spectrum with temperature of $T \ = \ 8122 \ \pm \ 273$ K, which is comparable with the results of previous studies that reported the spectral energy distribution of near-ultraviolet to optical during the flare could be approximated by single-temperature blackbody radiation at $T \ \sim \ 10^{4}$ K. We also estimated the time evolution of the flare temperature during the decay phase, which has not been reported with high-time-cadence spectroscopy. The radiative energy of this flare within the optical range is $4.4 \ \times \ 10^{32}$ erg, taking into account the time-dependent variation in the decreasing flare temperature and expanding flare area. Furthermore, we detected a delayed increase in the flux of H$\alpha$ after the photometric flare peak, secondary increase, and gradual increase even after the white-light flare ended. Comparison of our results with light curves obtained by the Sun-as-a-star analysis of solar flares indicates that these signals may be due to postflare loops near the stellar limb. Our result about time evolution of white-light continuum will help to gain more insight into the mechanism of white-light flares both on the Sun and stars. Additionally, since extreme ultraviolet radiation from flare loops plays a key role in planetary atmospheric escape, the existence of postflare loops on stellar flares and its time evolution will help future studies about habitability of close-in planets. 
\end{abstract}

\section{Introduction}
The Sun suddenly releases accumulated magnetic energy around sunspots, which is known as solar flares. The trigger of flares is thought to be magnetic reconnection in the solar corona (e.g., \cite{Shibata and Magara 2011}; 
\cite{Masuda et al. 1994}). Solar and stellar flares are observed across a wide wavelength range from radio to gamma-rays. Not only on the Sun but also on various types of stars, sudden increase of luminosity happen across multi-wavelength, including X-rays, ultraviolet (UV), optical, and infrared (IR), which is known as stellar flares (e.g., \cite{Kowalski 2024}; \cite{Benz 2017}).

The energy released from solar flares is typically $10^{29}-10^{32}$ erg (e.g., \cite{Shibata and Magara 2011}). In the case of stellar flares, previous studies have found a lot of superflares, whose energy is more than 10 times the energy of the largest solar flares (e.g., \cite{Maehara et al. 2012}; \cite{Notsu et al. 2019}; \cite{Okamoto et al. 2021}). In recent years, spectroscopic observations have been conducted to explore the physical mechanisms of flares and associated plasma ejections (\cite{Honda et al. 2018}; \cite{Maehara et al. 2021}; \cite{Namekata et al. 2022a}; \cite{Namekata et al. 2022b}; \cite{Notsu et al. 2024}; \cite{Namekata et al. 2024}). Among these studies, some papers have reported similarities in multiwavelength radiation between solar and stellar flares. \citet{Maehara et al. 2015} reported that the flare frequency distributions of superflares on Sun-like stars and solar flares are roughly on the same power-law line for the wide energy range between $10^{24}$ and $10^{35}$ erg. \citet{Neupert 1968} found an empirical relationship in solar flares, known as the Neupert effect, in which the time evolution of soft X-ray emission corresponds to the time-integrated hard X-ray emission. \citet{Hawley et al. 1995} reported a stellar Neupert effect, which implies that the time-integrated white-light radiation corresponds to the coronal radiation (see also \cite{Mitra-Kraev et al. 2005}; \cite{Tristan et al. 2023}). Shibata \& Yokoyama (\yearcite{Shibata and Yokoyama 1999}, \yearcite{Shibata and Yokoyama 2002}) suggested a theoretical scaling law which describes a universal correlation between the temperature and emission measure among not only solar flares but also stellar flares.
The presence of these features suggests that solar and stellar flares occur via the same physical processes.

The majority of stellar flares show an increase in the continuum from near-ultraviolet (NUV) to optical wavelengths. These are thought to be stellar analogues of white-light flares on the Sun \citep{Kowalski 2024}. While white-light flares are rare events on the Sun, they often occur on other stars, especially late-type stars, because these stars exhibit stronger magnetic activity than the Sun and their background emissions are relatively weak (\cite{Fuhrmeister et al. 2008}). Since white-light continuum emission is spatially and temporally correlated with hard X-ray emission (\cite{Hudson et al. 2008}), its source is thought to be non-thermal electrons penetrating into the lower chromosphere or upper photosphere. 

The white-light continuum is often described by single-temperature blackbody radiation at $T \ \sim \ 10^4$ K (\cite{Hawley and Fisher 1992}). \citet{Kowalski et al. 2013} reported that white-light continua of various flares on M-dwarfs in the wavelength range of $4000-4800$ \AA\, were fitted with blackbody spectra at $T \ \sim \ 9000-14000$ K. \citet{Fuhrmeister et al. 2008} observed a white-light flare on the active M-dwarf CN Leo and found that white-light continuum during impulsive phase was well fitted with blackbody spectra at $T \ \sim \ 10000-15000$ K. \citet{Rabello Soares et al. 2022} observed more than 200 flares on F-K type stars with two color photometry, and they estimated the blackbody temperature and flaring area of white-light continuum. Although these papers reported the blackbody temperature at the flare peak and during the decay phase, the time evolution of flare temperature was not mentioned. \citet{Howard et al. 2020} observed flares on K5-M5 dwarfs using two photometric instruments, which are Evryscope and TESS, and estimated the flare temperatures. They reported that flare temperature decreased rapidly within several minutes after the flare peak. \citet{Bicz et al. 2025} conducted multi-color photometric observations, including TESS, on M- and K-dwarf flares. They also estimated time evolution of flare temperature and reported the rapid decay of flare temperature. However, blackbody radiation at high temperature has very small amounts of energy in the TESS bandpass, which makes a temperature determination increasingly difficult. It is also difficult to separate the time variation of line emissions from that of the white-light continuum using photometric observations alone. To examine the mechanism of white-light flares, it is necessary to investigate the temperature of flare component and its time evolution. Although there are a lot of reports about flare observations with TESS including \citet{Howard et al. 2020}, the number of high-time-cadence spectroscopic observations simultaneously with TESS is still small. 

Researches about stellar flares on M-dwarfs are also important in terms of close-in habitable exoplanets. If we reveal the energy distribution of continuum radiation from optical to NUV or far-ultraviolet (FUV) during white-light flares, this provides insight into the impact of stellar flares on photodissociation reactions of key atmospheric species (e.g., ozone, oxygen, water vapor, and methane) in exoplanetary environments(\cite{Airapetian et al. 2016}). In addition to that, EUV radiation is a key factor in driving atmospheric escape processes on exoplanets. \citet{Liu et al. 2024} reported that UV emissions from flare loops during the late phase of solar flares have a significant impact on the Earth's ionosphere. The presence of flare loops and understanding these temporal evolutions provide insights into how intense and how long EUV radiation associated with stellar flares persists. 

Radiation from post-flare loops must be taken into account when considering flare emission, and they are expected to be observed for stellar flares(\cite{Heinzel and Shibata 2018}; \cite{Yang et al. 2023}; Yudovich, Yang, and Sun \yearcite{Yudovich et al. 2025}). As possible signatures of stellar postflare loops, \citet{Namizaki et al. 2023} reported that the ratio of the equivalent width (EW) of the red-shifted emission component to that of the line center component increased during the late phase, which could indicate contributions from downflows along postflare loops. However, in observations of stellar flares, it is still difficult to separate the contributions from flare ribbons and postflare loops. By applying recent results from the Sun-as-a-star analyses (e.g., \cite{Otsu et al. 2024}), it may be possible to separate emission from post-flare loops in stellar flares and flare emission itself ,and thus detect postflare loops. To investigate postflare loops of stellar flares in great detail, we conducted high-time-cadence spectroscopic observations and observed temporal evolution of chromospheric lines.

To investigate the mechanisms of white-light flares and time evolution of chromospheric lines during white-light flares, we conducted time-resolved photometric and spectroscopic observations of the active M-dwarf EV Lacertae (EV Lac). In this paper, we report the results of our analysis of a white-light flare on EV Lac. Section 2 describes the details of our target star and observations. Section 3 presents the analysis and results of the observed flare. In section 4, we discuss the energy of the white-light flare with and without considering the time variation of flare temperature, and the indication of postflare loops by comparing the H$\alpha$ light curve with that of an X-class solar flare.

\section{Observations and data reduction}\label{sec:2}
\subsection{Target star : EV Lac}\label{ssec:21}
In this work, our target is M3.5-type dwarf EV Lac (GJ 873, HLS 3853). EV Lac has been the subject of numerous flare studies on M dwarfs because it is known as a highly active flaring star. For example, \citet{Schmidt et al. 2012} reported that its flare rate of photometry observations within the energy from $10^{31}$ to $10^{33}$ erg is $\sim$0.4 events per hour and that from infrared (IR) observations is $\sim$0.13 events per hour. Many flares have been detected in multi-wavelength, including NUV and optical \citep{Kowalski et al. 2013}, X-ray \citep{Inoue et al. 2024}, and IR \citep{Schmidt et al. 2012}. \citet{Osten et al. 2005} conducted multi-wavelength observations on EV Lac from radio to X-ray. The stellar parameters of EV Lac are given in Table 1.
\begin{table}[tb]
    \centering
    \label{tb:parameter}
    \tbl{Parameters of EV Lac}{
    \begin{tabular}{c c c}
    \hline
    \hline
    Parameters & Value & Reference\footnotemark[$*$]\\
    \hline
    Effective Temperature $T_{\rm{eff}}$ (K) & $3270 \ \pm \ 80$ & (1) \\
    Distance $d$ (pc) & $5.049 \ \pm \ 0.001$ & (1)\\
    Age (Myr) & 128-800 & (1)\\
    Stellar Radius $R_{\rm{star}}$ ($R_{\solar}$) & $0.331 \ \pm \ 0.013$ & (2) \\
    Stellar Mass $M_{\rm{star}}$ ($M_{\solar}$) & $0.320 \ \pm \ 0.008 $ & (2) \\
    Rotational Period $P_{\rm{rot}}$ (day)  &  $4.297 \ \pm \ 0.017$ & (3) \\
    Rotational Velocity $v_{\rm{rot}}\  \mathrm{({km \ s^{-1})}}$  & $4.2 \  \pm \ 0.5$ & (4) \\
    \hline
    \end{tabular}}
    
    \begin{tabnote}
        \footnotemark[$*$] (1)\citet{Paudel et al. 2021}; (2)\citet{Pineda et al. 2021}; (3)\citet{Muheki et al. 2020}; (4)\citet{Pettersen. 1980}
    \end{tabnote}
\end{table}

\subsection{Photometric data : TESS}\label{ssec:22}
The Transit Exoplanetary Survey Satellite (TESS; \cite{Ricker et al. 2015}) is a space telescope developed for NASA's Explorer program, designed to search for exoplanets using the transit method. TESS covers a wavelength range from 6000 to 10000 \AA. We used the Simple Aperture Photometry (SAP) light curve obtained from the Mikulski Archive for Spase Telescope (MAST) portal. In this work, we analyzed the two-min cadence data of EV Lac (TIC 154101678) in the Sector 16 (from 2019 September 11 to 2019 October 19).

\subsection{Spectroscopic data : Seimei telescope}\label{ssec:23}
We obtained spectroscopic data on 2019 September 14 with the Kyoto-Okayama Optical Low-dispersion Spectrograph Integral Field Unit (KOOLS-IFU; \cite{Matsubayashi et al. 2019}) on the 3.8 m Seimei telescope \citep{Kurita et al. 2020} at Okayama Observatory. We employed the low-resolution VPH-blue grism, which provides continuous wavelength coverage from 4100 to 8900 \AA, with a spectral resolution of $\lambda/\Delta\lambda \ \sim \ 500$. We can obtain two-dimentional spectroscopic data with this IFU instrument. The field of view (FoV) of one fiber is regular hexagon of $0.93 \ \pm \ 0.04$ arcsec in diameter. The total FoV is $15.1 \ \pm \ 0.7$ arcsec in diameter. We obtained the flux only from the target star because of this small FoV. By integrating spectra of each fiber, this instrument enables us to conduct photometric and spectroscopic observations. Since this is not relative photometry, it is noted that observational data may be affected by systematics such as elevation during long observations.  

The exposure time was set to 30 seconds, and readout time was $\sim22$ seconds, resulting in a total cadence of $\sim1$ minute, which is shorter than that of TESS photometric data from Sector 16. 

We conducted data reduction by following the methods outlined in \citet{Namekata et al. 2020}, using the IRAF\footnotemark[1]\footnotetext[1]{ IRAF and PyRAF are distributed by the National Optical Astronomy Observatories, which are operated by the Associationo fUniversities for Research in Astronomy, Inc., under cooperative agreement with the National Science Foundation} package, PyRAF\footnotemark[1] software, the data reduction packages developed by \citet{Matsubayashi et al. 2019}\footnotemark[2]\footnotetext[2]{http://www.o.kwasan.kyoto-u.ac.jp/inst/p-kools/reduction-201806/index.html}. The reduction processes included overscan correction, bias subtraction, gain correction, flat fielding, wavelength calibration, sky subtraction, fiber integration, and flux calibration. We performed flat fielding, wavelength calibration, and flux calibration using the twodspec package in IRAF. For flux calibration, we also observed HR7596 as a standard star. Since the position of the target star was changing during observation due to tracking error and atmospheric dispersion, we integrated spectra of all 127 fibers after performing sky subtraction to minimize the flux loss due to the changes in PSF size during our observation. However, there is a possibility of flux loss around the beginning and the end of the observation because the elevation of the target star is low at those times. The elevation and the airmass during Seimei observation is shown in appendix 1.

\section{Analysis and results}\label{sec:3}
\subsection{Observed flare}
\begin{figure*}[h]
 \begin{center}
  \includegraphics[width=16cm]{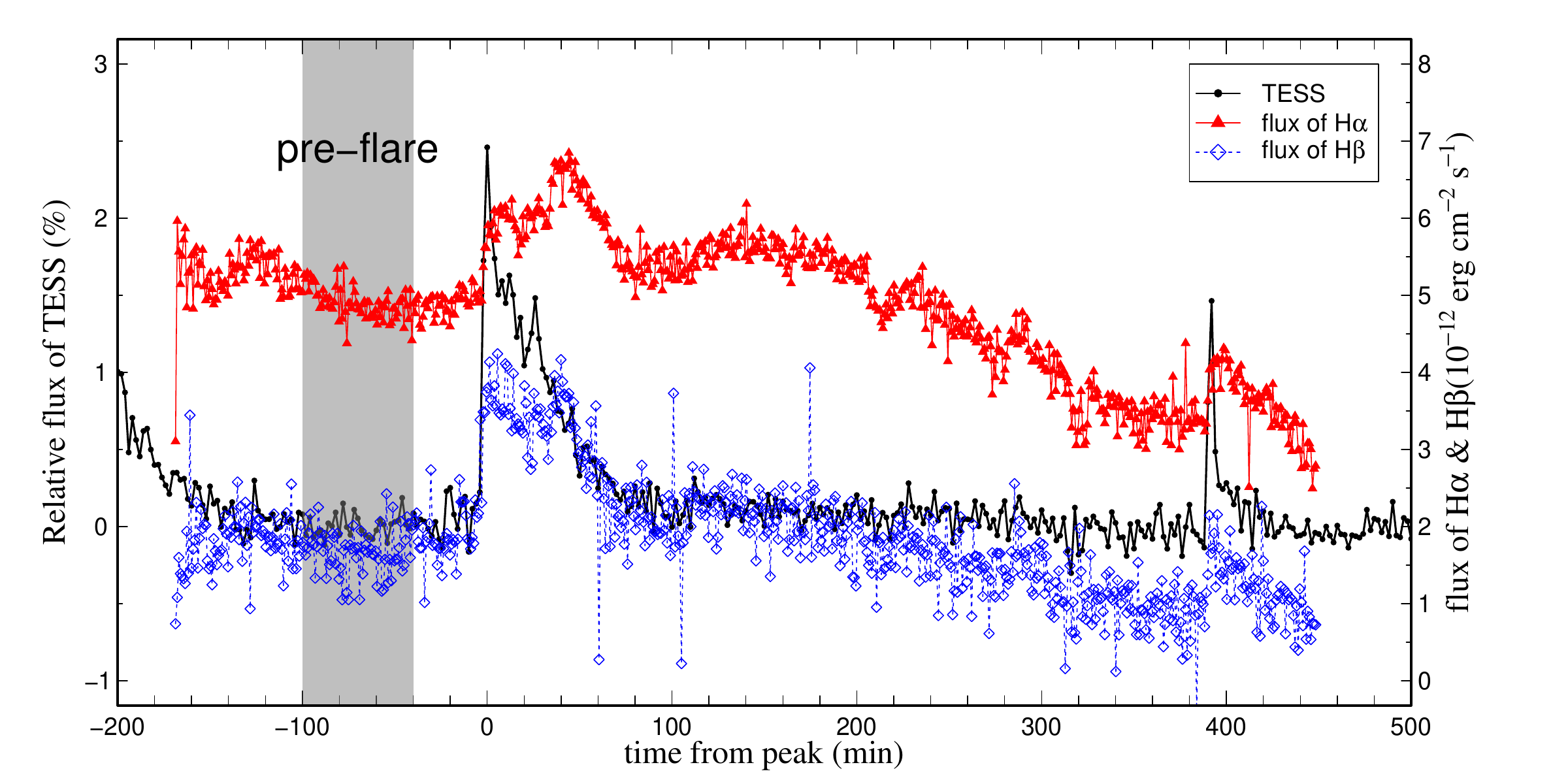} 
 \end{center}
\caption{ Light curves of EV Lac of the 2-minute cadence TESS and the line fluxes of H$\alpha$ and H$\beta$ during the whole Seimei observation. The long-term trend of the TESS light curve has already been removed. The horizontal axis represents the time from the flare peak in minutes. The left-hand vertical axis shows the relative flux of TESS (\%) normalized to the average flux during the pre-flare quiescent state. The gray shaded area, spanning the -100 to -40 minutes from the flare peak, marks the pre-flare quiescent state in this study. The right-hand vertical axes indicate the line flux of H$\alpha$ and H$\beta$, respectively. The filled circles, red triangles, and open blue diamonds represent the flux of TESS photometry, the line flux of H$\alpha$, and H$\beta$, respectively. 
{Alt text: Line graph. The x axis represents the time from the flare peak from -200 to 500 minutes. The y axis shows raw counts of TESS from 1.10 to 1.20 times ten to the fifth. The right-hand y axis shows the line fluxes of H alpha and H beta from 0 to 8 times ten to the minus twelfth erg per square centimeters and per second.} 
}\label{fig:obs_alltime}
\end{figure*}

We observed a flare on 2019 September 14 UT (Barycentric Julian Date (BJD): 2458741.03) with both the Seimei telescope and TESS simultaneously. Figure \ref{fig:obs_alltime} shows the light curve of EV Lac observed by TESS and Seimei during whole Seimei observation. The long-term modulation of the TESS light curve has already been removed. The method of detrending and detrended light curve are in appendix 2. 

\begin{figure}[tb]
 \begin{center}
  \includegraphics[width=8cm]{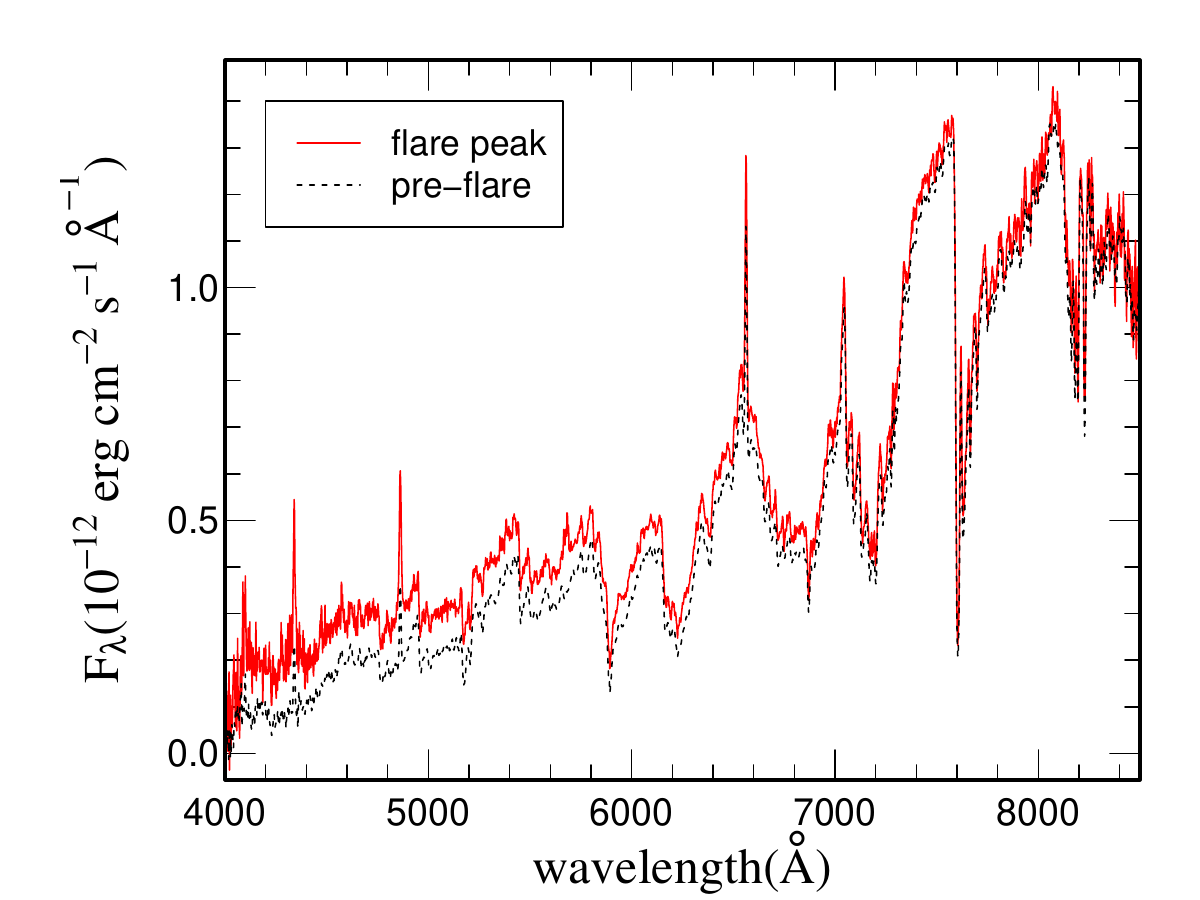} 
 \end{center}
\caption{ Two Spectra of EV Lac obtained with the Seimei telescope. The red solid line and black dashed line represent the spectra at the flare peak and at the pre-flare quiescent state, respectively. The pre-flare spectrum was created by averaging the spectra during the pre-flare quiescent state (see gray shaded area in figure \ref{fig:obs_alltime}).
 {Alt text: : Line graph. The x axis shows wavelength from 4000 to 8500 angstrom. The y axis shows the flux density from 0 to 1.5 times ten to the minus twelfth erg per square centimeters and per second and per angstrom. } 
}\label{fig:pre_and_peak}
\end{figure}

\begin{figure*}[ht]
 \begin{center}
  \includegraphics[width=12cm]{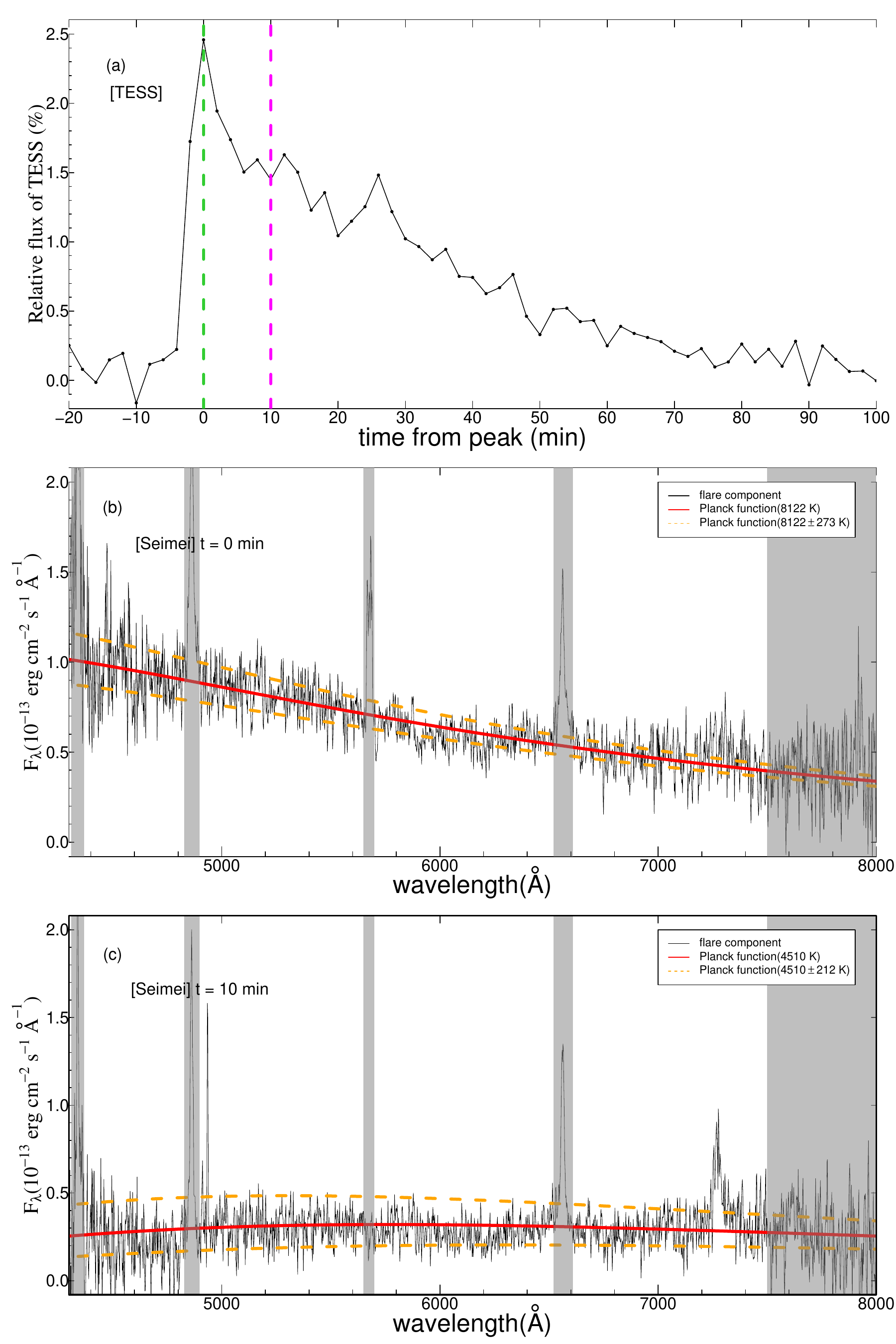} 
 \end{center}
\caption{ (a) The vertical dashed line marks the time corresponding to the spectrum shown in panel (b) and (c). (b) The spectrum of flare component at t = 0 min. The red thick curve indicates the best fit scaled Planck function with a temperature of $T_{\mathrm{BB}} \ = \ 8122$ K. The upper- and lower-dashed curves indicate the Planck function considering the error of temperature at $ \ 8122 \ \pm \ 273$ K and the error of the amplitude $A$, respectively. 
The gray shaded area indicates the data which we did not use for fitting. (c) Same as panel (b), but the spectrum is at $t = 10$ min. The horizontal and vertical axes are same as panel (b). Red thick curve indicates the best fit scaled Planck function with a temperature of $T_{\mathrm{BB}} \ = \ 4510$ K, while upper- and lower-dashed curves indicate scaled Planck functions considering the error of temperature at $ \ 4510 \ \pm \ 212$ K and the error of the amplitude $A$, respectively. 
{Alt text: : Three line graphs. In the upper panel, the x axes show time from the flare peak from -20 to 100 minutes. The y axes show relative flux from -0.1 to 2.5 percents. In the middle and lower panels, the x axes show wavelength from 4300 to 8000 angstrom. The y axes show flux density from 0 to 2.0 times ten to the minus thirteenth erg per square centimeter and per second and per angstrom. } 
}\label{fig:2spec_with_tess}
\end{figure*}

\begin{figure*}[ht]
 \begin{center}
  \includegraphics[width=13cm]{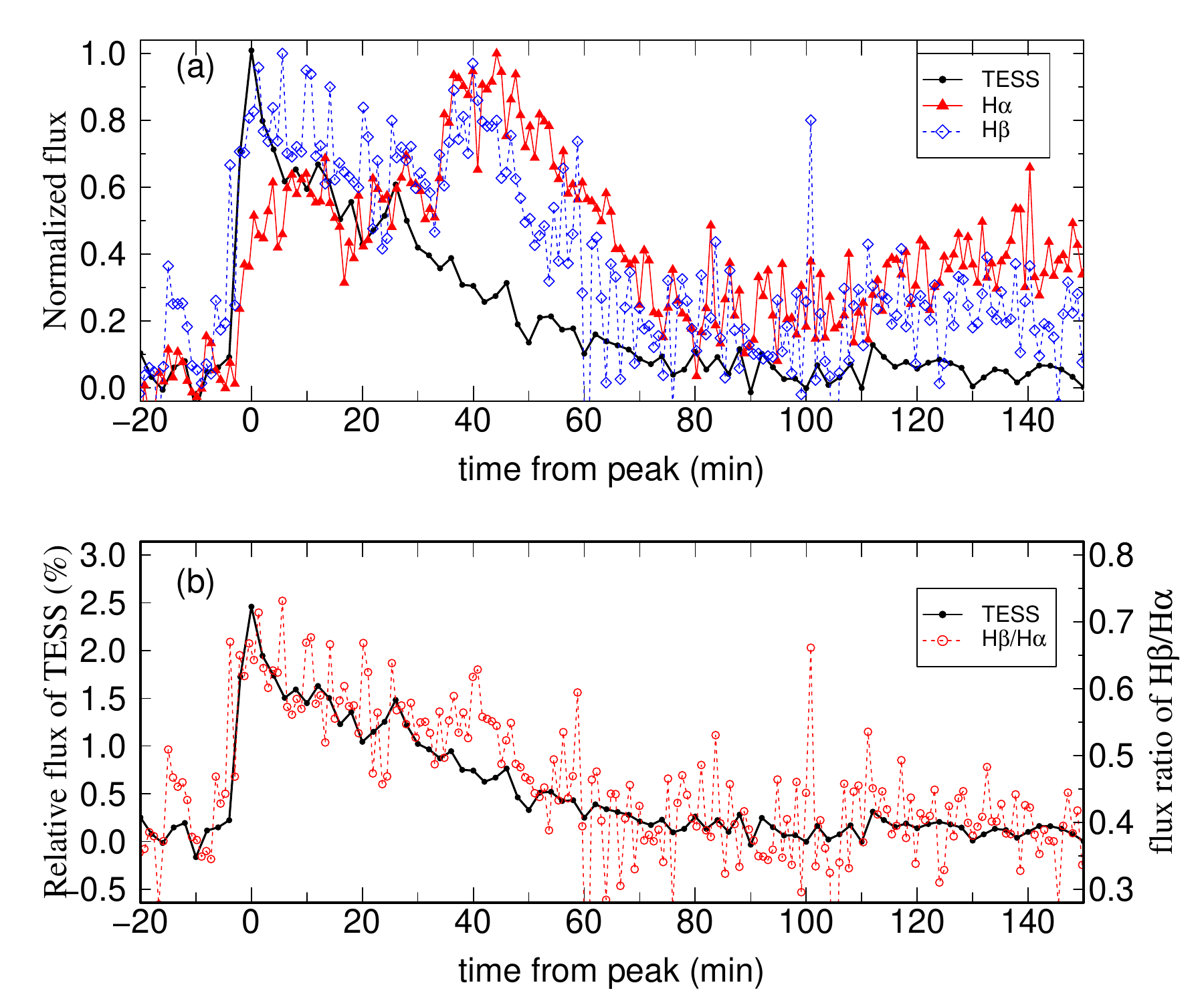} 
 \end{center}
\caption{ (a) Light curves of photometric flux and Balmer-lines flux. The vertical axis represents the normalized flux, and the horizontal axis represents time in minutes relative to the flare peak observed with TESS. The fluxes are normalized to their respective peak values. The colors of the lines and markers are the same as in figure \ref{fig:obs_alltime}. 
(b) Time variation of the photometric flux observed with TESS and the flux ratio of H$\beta$ to H$\alpha$. The horizontal axis represents the time from the flare peak in minute. The left-hand vertical axis shows the relative flux (\%) normalized to the average flux of the pre-flare quiescent state (same as in figure \ref{fig:obs_alltime}), while the right-hand vertical axis represents the flux ratio of H$\beta$/H$\alpha$. The filled circles and open red circles represent the flux of TESS photometry and the flux ratio of H$\beta$/H$\alpha$, respectively. 
{Alt text: : Two line graphs. In both panels, the x axes show time from the flare peak from -20 to 150 minutes. In the upper panel, the left-hand y axis shows normalized flux from 0 to 1.0. In the lower panel, the left- and right-hand y axes show relative flux from -0.5 to 3.0 percents, and the flux ratio of H beta to H alpha from 0.3 to 0.8, respectively.} 
}\label{fig:Balmer_flux_and_ratio}
\end{figure*}

Figure \ref{fig:pre_and_peak} presents two spectra of EV Lac observed with the Seimei telescope, one at the flare peak and the other as an average of the pre-flare quiescent state. 
By subtracting the pre-flare quiescent spectrum from the spectra obtained during the flare, we derived the spectra of the flare component. Figure \ref{fig:2spec_with_tess} shows two spectra of the flare component at the flare peak and 10 minutes after the peak. 
We also derived light curves of the H$\alpha$ ($6562.8$ \AA) and H$\beta$ ($4861.3$ \AA) lines. Figure \ref{fig:Balmer_flux_and_ratio}(a) shows the light curve of the relative flux of TESS, H$\alpha$, and H$\beta$, each normalized to their respective peak values. 
The line flux of H$\alpha$ and H$\beta$ are calculated with the wavelength ranges from $6550$ to $6576$ \AA \, and $4841.3$ and $4881.3$ \AA.

\subsection{Time variation of flare temperature}
\begin{figure}[htb]
 \begin{center}
  \includegraphics[width=8cm]{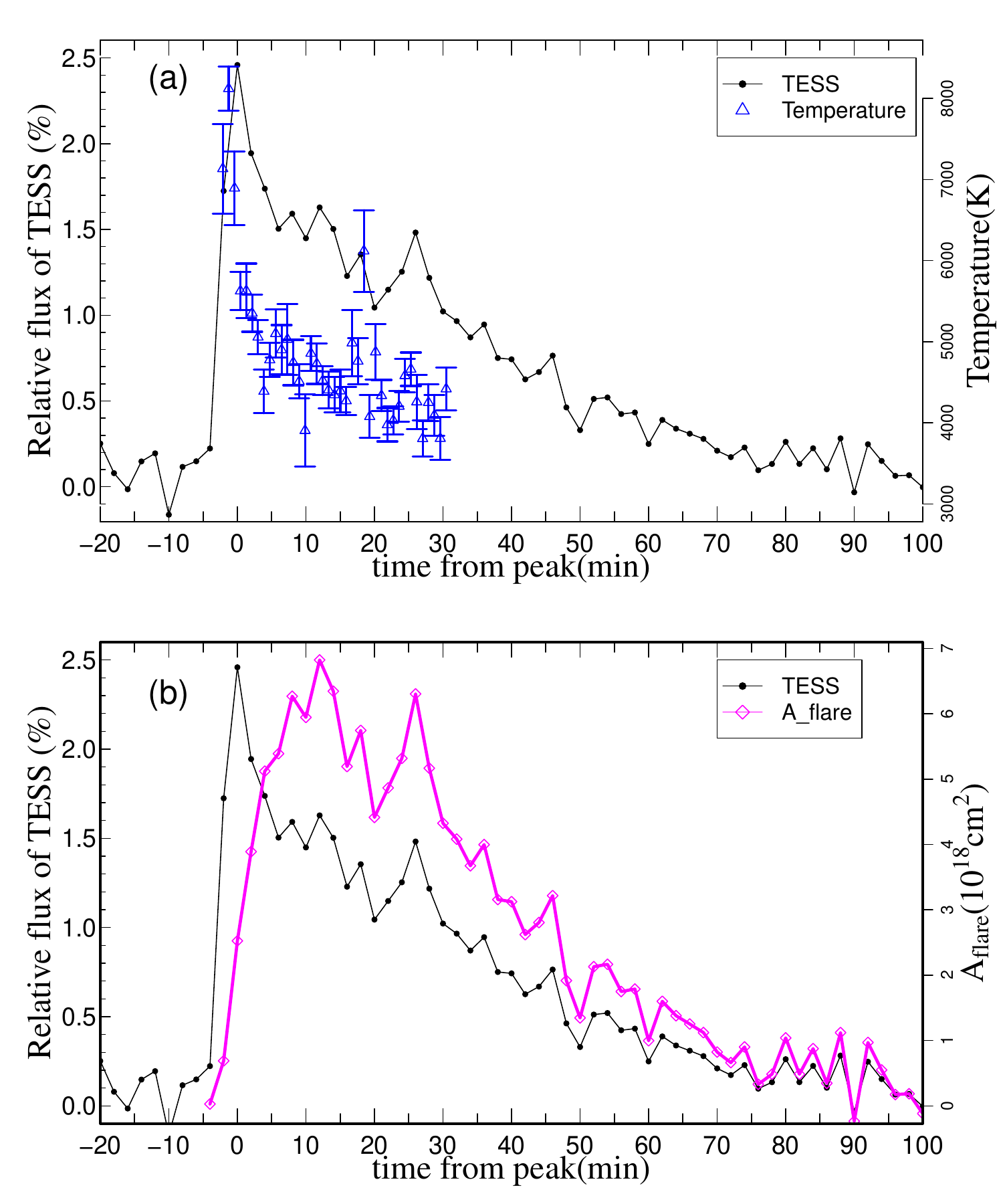} 
 \end{center}
\caption{ (a) Time variation of the photometric flux observed with TESS and the flare temperature $T_{\mathrm{BB}}$. The horizontal axis represents the time from the flare peak in minutes. The left-hand vertical axis represents the relative flux (\%) normalized by the average flux of pre-flare quiescent state (same as figure \ref{fig:obs_alltime}), while the right hand axis represents the flare temperature $T_{\mathrm{BB}}$ (K). Filled black dots indicate the photometric light curve of TESS. Open triangles indicate the flare temperatures derived from blackbody fitting, with error bars representing $3\sigma_{\mathrm{T}}$. (b) Time variation of photometric flux observed with TESS and flare area $A_{\mathrm{flare}}$. The horizontal axis represents the time from the flare peak in minutes. The left-hand vertical axis shows the relative flux (\%) normalized by the average flux of the pre-flare quiescent state (same as in figure \ref{fig:obs_alltime}), while the right-hand vertical axis represents the flare area ($\times \ 10^{18} \ \mathrm{cm}^{2}$). Filled black dots and open diamonds indicate the TESS photometric light curve and the flare area $A_{\mathrm{flare}}$, respectively.
{Alt text: Two line graphs and dot plot. The x axes show time from the flare peak, ranging from -20 to 100 minutes. In the both panel, the left-hand y axes show relative flux from -0.2 to 2.6 percents. In the upper panel, the right-hand y axes show the temperature from 3000 to 8500 kelvin. In the lower panel, the right-hand y axis shows the flare area from 0 to 7.0 times ten to the eighteenth square centimeters.} 
}\label{fig:temperature_and_area}
\end{figure}

First, we estimate the temperature of the flare component from flare spectra obtained by the Seimei telescope. As mentioned in section 1, many studies on white-light flares in late-type stars have reported that flare emission from NUV to optical range can be approximated as single-temperature blackbody radiation at $T_{\mathrm{BB}} \ \sim \ 10^{4}$ K (\cite{Hawley and Fisher 1992}; \cite{Kowalski et al. 2013}). We estimate the blackbody temperature of the flare components by fitting a scaled Planck function to the flare spectra. To perform this estimation, we use the \texttt{optimize.curve\_fit} module from \texttt{SciPy} package. The scaled Planck function is given by 
\begin{eqnarray}
 B_{\lambda} = \frac{A}{\lambda^{5}}\frac{1}{\exp\left(\frac{hc}{\lambda k_{\mathrm{B}} T_{\mathrm{BB}}}\right) \ - \ 1} ,
 \label{Planck function}
\end{eqnarray}
where $A$ is a scale factor, $\lambda$ is the wavelength, $h$ is Planck constant, $c$ is light speed, $k_{\mathrm{B}}$ is Boltzmann constant, and $T_{\mathrm{BB}}$ is the blackbody temperature, respectively. $A$ and $T_{\mathrm{BB}}$ are the free parameters. To avoid some emission lines, the wavelength range in which we used the data for blackbody fitting is $4300 \ < \ \lambda \ < \ 7500$ \AA \,except for Balmer lines and outlier. To be specific, we used $4300 \ < \ \lambda \ < \ 4310$ \AA, $4370 \ < \ \lambda \ < \ 4828$ \AA, $4898 \ < \ \lambda \ < \ 5650$ \AA, $5700 \ < \ \lambda \ < \ 6520$ \AA, and $6610 \ < \ \lambda \ < \ 7500$ \AA. As a result of this, we obtained the flare temperature during the flare, as shown in figure \ref{fig:temperature_and_area}(a). In particular, the temperature and the scale factor at the flare peak are $T_{\mathrm{BB}} \ = \ 8122 \ (\pm \ 273)$ K and $7.586 \ (\pm \ 0.888) \ \times \ 10^{-29}$. We obtained the standard deviation of temperature ($\sigma_{\mathrm{T}}$) and scale factor ($\sigma_{\mathrm{A}}$) 
from the covariance matrix that the \texttt{scipy.curve\_fit} routine returns. In this work, we use credible region $3\sigma_{\mathrm{T}}$ and $3\sigma_{\mathrm{A}}$ as the error of temperature and scale factor. Figure \ref{fig:2spec_with_tess}(b) presents the spectrum of the flare component at the flare peak along with the scaled Planck functions corresponding to $T_{\mathrm{BB}} \ = 8122 \ \pm \ 273$ K. 

Then, we investigated the time evolution of the flare temperature during the flare. Some previous studies investigated the time variation of the flare temperature. \citet{Hawley et al. 2003} observed flares on a M-dwarf AD Leonis (AD Leo) and estimated flare temperature within a few minutes from the beginning of the flare by using UV spectrum and \textit{UBVR}-band photometric flux. Seeing figure 10 in \citet{Hawley et al. 2003}, in general the blackbody continuum does not fit all of the fluxes within the errors, but it provides the correct overall shape of the blackbody continuum. From figure 11 in \citet{Hawley et al. 2003}, the flare temperature is about 8000-10000 K within a few minutes from the beginning of the flare. \citet{Howard et al. 2020} estimated the flare temperature on M-dwarfs from the spectral energy distribution (SED) analysis, incorporating photometric data from TESS and Evryscope. \citet{Howard et al. 2020} reported that the flare temperature decreased dramatically (within several minutes) after the flare peak. \citet{Bicz et al. 2024} observed the long-duration flare on magnetically active young star CD-36 3202 with TESS. They also investigated the time evolution of flare temperature using the modified method from \citet{Shibayama et al. 2013}. First, they calculated the size of the flaring region at the flare peak with fixed temperature $T_{\mathrm{flare}} \ \sim \ 10^4$ K. Then, assuming that the size of flaring region is constant during the whole flare, they calculated the flare temperature (see figure 4 in \cite{Bicz et al. 2024}). The flare temperature of the flare peak was $\sim \ 12000$ K in \citet{Bicz et al. 2024}. 

We examined the time evolution of the flare temperature using time-resolved spectroscopic data. For instance, figure \ref{fig:2spec_with_tess}(c) shows the spectrum of flare component 10 minutes after the flare peak along with the best-fitted Planck functions (at $T_{\mathrm{BB}} \ = \ 4510 \pm \ 212$ K). Notably, the flare temperature decreased by almost half within just 10 minutes. The time variation of flare temperature is shown in Figure \ref{fig:temperature_and_area}(a). Figure \ref{fig:temperature_and_area}(a) reveals that the decay of the flare temperature is significantly more rapid than the decay of the photometric flux observed with TESS. We calculated the e-folding time of decaying temperature and photometric flux by fitting exponential functions to the decay phase. As a result, the e-folding time of the temperature was found to be $\sim \ 2.5$ min, whereas that of the photometric flux was $\sim \ 30$ min. The e-folding time of the temperature is more than 10 times shorter than that of the white-light flux. Compared with figure 5 of \citet{Howard et al. 2020}, this result is in general consistent with the rapid decay of the flare temperature reported in their paper. Figure 4 in \citet{Bicz et al. 2024} also showed the time variation of flare temperature. In their paper, the time variation of the flare temperature was similar to the light curve of TESS (figure 1(a) in \cite{Bicz et al. 2024}), but the duration of the flare and the method of estimating the temperature were different from our analysis. For future work, we have to investigate the dependence of time variation of flare temperature on duration and energy of flares.

\subsection{Energy estimation}
We calculated the flare energy following the method in \citet{Shibayama et al. 2013}. Using equations (1)-(6) from \citet{Shibayama et al. 2013}, we estimated the total flare energy based on stellar luminosity, flare amplitude, flare duration, and the flare temperature. 

First, we estimated the flare area ($A_{\mathrm{flare}}(t)$) using the observed luminosity of the star in its quiescent state ($L'_{\mathrm{star}}$), the luminosity during the flare ($L'_{\mathrm{flare}}(t)$), and the flare amplitude of the light curve obtained with TESS ($C'_{\mathrm{flare}}(t)$). These quantities are defined by the following equations:

\begin{eqnarray}
 L'_{\mathrm{star}} = \int R_{\lambda}B_{\lambda(T_{\mathrm{eff}})}d\lambda \cdot \pi R_{\mathrm{star}}^2,
 \label{L'_star}
\end{eqnarray}

\begin{eqnarray}
 L'_{\mathrm{flare}}(t) = \int R_{\lambda}B_{\lambda(T_{\mathrm{BB}}(t))}d\lambda \cdot A_{\mathrm{flare}}(t), \ \mathrm{and}
 \label{L'_star}
\end{eqnarray}

\begin{eqnarray}
 C'_{\mathrm{flare}}(t) = \frac{L'_{\mathrm{flare}}(t)}{L'_{\mathrm{star}}},
 \label{Planck function}
\end{eqnarray}
\noindent
where $T_{\mathrm{BB}}(t)$ is the flare temperature, $R_{\lambda}$ is the response function of TESS (see figure 1 in \cite{Ricker et al. 2015}), $B_{\lambda(T)}$ is the Planck function, and $R_{\mathrm{star}}$ and $T_{\mathrm{eff}}$ are the radius and efficient temperature of the target star, respectively (see table 1). The spectral bandpass of TESS ranges approximately from 6000 \AA to 10000 \AA. Using equations (2)-(4), we estimated $A_{\mathrm{flare}}(t)$ as follows,

\begin{eqnarray}
 A_{\mathrm{flare}}(t) = C'_{\mathrm{flare}}(t)\pi R_{\mathrm{star}}^2\frac{\int R_{\lambda}B_{\lambda(T_{\mathrm{eff}})}d\lambda}{\int R_{\lambda}B_{\lambda(T_{\mathrm{BB}}(t))}d\lambda}.
\end{eqnarray}

Assuming that the flare component behaves as a blackbody radiator, we calculated the luminosity of the flare within the TESS bandpass ($L_{\mathrm{flare}}(t)$) using $T_{\mathrm{BB}}(t)$, $A_{\mathrm{flare}}(t)$, and Stefan-Boltszmann constant ($\sigma_{\mathrm{SB}}$),

\begin{eqnarray}
 L_{\mathrm{flare}}(t) = \sigma_{\mathrm{SB}}T_{\mathrm{BB}}(t)^{4}A_{\mathrm{flare}}(t).
 \label{L_flare}
\end{eqnarray}

$T_{\mathrm{BB}}(t)$ is a time-dependent parameter, whereas the corresponding parameter used in \citet{Shibayama et al. 2013} was treated as time-independent. Instead of directly using $T_{\mathrm{BB}}(t)$, we applied an exponential function fitted to $T_{\mathrm{BB}}(t)$ obtained from flare spectra in Section 3.2. The fitting was performed over the range where the relative flux observed by TESS was greater than 1 \% ($> \ 1$ \%), as $T_{\mathrm{BB}}(t)$ exhibited large uncertainties when the relative flux was below this threshold ($< \ 1$ \%). Using $A_{\mathrm{flare}}(t)$ from equation (5), we computed $L_{\mathrm{flare}}(t)$ in equation (6). Finally, the total flare energy ($E_{\mathrm{flare}}$) was determined by integrating $L_{\mathrm{flare}}(t)$ over the duration of the flare,

\begin{equation}
 E_{\mathrm{flare}} = \int^{t_{\mathrm{end}}}_{t_{\mathrm{start}}}L_{\mathrm{flare}}(t)dt.
\end{equation}

We integrated $L_{\mathrm{flare}}(t)$ from $t_{\mathrm{start}} \ = \ 2458741.027$ to $t_{\mathrm{end}} \ = \ 2458741.1$ in BJD, obtaining a total flare energy of $E_{\mathrm{flare}} \ = \ 4.4 \ \times \ 10^{32} \ \mathrm{erg}$. 

Additionally, we estimated the flare energy using equations (2)-(7), assuming two different blackbody temperatures that were fixed during the flare: $T_{\mathrm{BB}} \ = \ 8122$ K (the peak flare temperature for this event) and $T_{\mathrm{BB}} \ = \ 10000$ K (a commonly used flare temperature in stellar flare studies). The calculated energy with a temperature of $T_{\mathrm{BB}} \ = \ 8122$ K was $8.9 \ \times \ 10^{32}$ erg, while at $T_{\mathrm{BB}} \ = \ 10000$ K, it was $1.2 \ \times \ 10^{33}$ erg.

\section{Discussion}
\subsection{Flare energy}
Several studies on stellar flares have estimated the flare energy from the NUV to optical range using the equations in \citet{Shibayama et al. 2013} (e.g., \cite{Gunther et al. 2020}; \cite{Namekata et al. 2017}; \cite{Yang and Liu 2019}). In this work, we calculated the flare energy using the time-dependent parameter $T_{\mathrm{BB}}(t)$. In contrast, in almost all previous studies, the flare energy has not been estimated with time-dependent flare temperature obtained from time-resolved spectroscopic observations. Under the assumption that the flare emission follows blackbody radiation, its luminosity is proportional to $T_{\mathrm{BB}}^{4}$ and $A_{\mathrm{flare}}(t)$ is also the function of $T_{\mathrm{BB}}(t)$. As shown in subsection 3.2, $T_{\mathrm{BB}}(t)$ decayed rapidly after the flare peak. Few studies estimated the flare energy without considering the flare temperature that is changing during the flare. Thus, we have to check how different the flare energy is calculated with the time-dependent flare temperature $T_{\mathrm{BB}}(t)$.

In subsection 3.3, we calculated the flare energy both by considering the time variation of the flare temperature ($T_{\mathrm{BB}}(t)$) and by assuming that $T_{\mathrm{BB}}$ remains constant during the flare. The resulting energy values calculated with fixed flare temperature ($E_{\mathrm{flare}} \ \sim \ 8.9 \ \times \ 10^{32} \ \mathrm{erg}$ for $T_{\mathrm{BB}} \ = \ 8122$ K, and $E_{\mathrm{flare}} \ \sim \ 1.2 \ \times 10^{33} \ \mathrm{erg}$ for $T_{\mathrm{BB}} \ = \ 10000$ K) were about 2-3 times greater than the energy estimated when accounting for the time variation of the flare temperature ($E_{\mathrm{flare}} \ \sim \ 4.4 \ \times \ 10^{32}$ erg). According to \citet{Schmitt et al. 2019}, when we assume the blackbody radiator, the ratio of flare energy within TESS-band (about $6000-10000$ \AA) to bolometric flare energy exhibits a temperature dependence (see figure 12 in \cite{Schmitt et al. 2019}). If the time variation of flare temperature is ignored, the energy can be overestimated by a factor of 2 to 3.

We also estimated the flare area of white-light ($A_{\mathrm{flare}}(t)$). Figure \ref{fig:temperature_and_area}(b) shows the time variation of $A_{\mathrm{flare}}(t)$ along with the photometric flux obtained with TESS. In almost all studies, the time evolution of the area of white-light flares ($A_{\mathrm{flare}}$) has not been investigated with optical spectra obtained from high-time-cadence spectroscopic observations. At the flare peak, $A_{\mathrm{flare}}(t)$ is still relatively small. However, during the decay phase, it expands and reaches its maximum at $t \ = \ 14$ min. After peaking, $A_{\mathrm{flare}}(t)$ gradually decreases. 
The maximum value of $A_{\mathrm{flare}}(t)$ was $\sim \ 6.8 \ \times \ 10^{18} \ \mathrm{cm}^2$, corresponding to about 0.3 \% of the disk area of EV Lac. \citet{Ikuta et al. 2023} conducted mapping starspots from the TESS light curve for two and three spots, and reported that the size of these starspots are about 5 \% of the disc area of EV Lac, corresponding to about 16 times larger than the maximum value of $A_{\mathrm{flare}}(t)$. It is noted that $A_{\mathrm{flare}}$ represents the white-light emitting region, not the flare ribbons.  

\citet{Hawley et al. 2003} investigated the time evolution of not only the flare temperature but also the fractional flare area coverage, which corresponds to $A_{\mathrm{flare}}$ (see figure 11 in \cite{Hawley et al. 2003}). They reported that the flaring area is about a few percent of visible hemisphere, which is almost the same as our results. \citet{Rabello Soares et al. 2022} observed 209 stellar flares on F-K type stars, and estimated the blackbody temperature and $A_{\mathrm{flare}}$ by using two-color photometry data. Their estimated $A_{\mathrm{flare}}$ ranged within $10^{18-23} \ \rm{cm}^2$. The $A_{\mathrm{flare}}$ of our observed flare is within the estimated values of \citet{Rabello Soares et al. 2022}. The top panel of figure 14 in \citet{Rabello Soares et al. 2022} shows that the Gaia flare energy and the flare area are correlated. Although their observations were on F-K type stars, not on M-type stars, our results ($E_{flare} \ \sim \ 4.4 \ \times \ 10^{32}$ erg and $A_{\mathrm{flare}} \ \sim \ 6.8 \ \times \ 10^{18} \ \mathrm{cm}^2$) almost follow this relation. 

As mentioned above, few previous studies examined the time evolution of white-light emitting region using time-resolved observations on stellar flares and solar flares. For future studies, in order to investigate the mechanism of white-light flares, more observations on white-light flares both on the Sun and stars is needed to compare white-light emitting region on solar flares with that on stellar flares. 

\subsection{Indications of Postflare loops based on evolution of Balmer line fluxes}
In this subsection, we discuss the time variation of H$\alpha$ and H$\beta$. In figure \ref{fig:Balmer_flux_and_ratio}(a), we observe that the first peak of the Balmer lines occurs $\sim \ 10$ minutes after the TESS peak. As we mentioned in section 1, white-light emission is thought to originate primarily from the heated chromosphere and/or photosphere. High-energy particles derived from magnetic reconnection penetrate chromosphere or the upper photosphere, where they heat the atmosphere. As a result, plasma in the chormosphere is heated very much, leading to the formation of chromospheric condensation \citep{Fisher et al. 1985} with a temperature of order of $\sim \ 10^4$ K, which can be a source of white light emission as well as the H$\alpha$ emitting downflow (origin of the red asymmetry, \cite{Ichimoto and Kurokawa 1984}). On the other hand, the temperature of the plasma above chromospheric condensation is heated to $10^7$ K, making upward flow known as chromospheric evaporation observed in soft X-rays and EUVs \citep{Nagai 1980}. Hence, the footpoint of this part (hot flare loop observed by soft X-rays and EUVs) is the main source of H$\alpha$ radiation (see review by \cite{Kowalski 2024}). This hot flare loop gradually cools down, from $10^7$ K (observed in soft X-rays) to $10^6$ K (in EUVs), eventually to $10^4$ K (in H$\alpha$, e.g., Kamio, Kurokawa, and Ishii \yearcite{Kamio et al. 2003}). This cool flare loop observed in H$\alpha$ is the post-flare loop we are discussing here.

Examining the light curve of H$\alpha$ flux in figure \ref{fig:Balmer_flux_and_ratio}(a),
we identify three key features: the first peak delayed relative to the flare peak of white-light ($t \ \sim \ 10$ min), a secondary peak ($t \ \sim \ 40$ min), and a gradual increase following the end of the white-light flare ($t \ \sim \ 100$ min). Although flare ribbons contribute to the first peak of H$\alpha$, it is not easy to understand the second peak and the gradual increase in H$\alpha$ with contribution from flare ribbons. We propose two possible interpretations for these features. 

The first is that non-white-light flares occurred during the decay phase. 
During the decay phase, there was a small increase in optical continuum flux (TESS) at $t \ \sim \ 25$ min. However, since this amplitude is much smaller compared to the main flare signal, we concluded that this signal was insignificant to define this increment as another white-light flare. 
Some previous studies on stellar flares have reported the existence of non-white-light flares. 
\citet{Maehara et al. 2021} reported an H$\alpha$ flare that was not accompanied by an increase in the white-light continuum or H$\beta$ (see flare C in figure 5(a) and (b) of \cite{Maehara et al. 2021}). \citet{Maehara et al. 2021} detected the increase in EW of H$\alpha$ after flare C. However, they concluded that this secondary increase in H$\alpha$ was another flare (flare D) because not only H$\alpha$ but also the EW of H$\beta$ and the white-light continuum flux increased simultaneously. They explained that the reason flare C did not appear in H$\beta$ was the lower optical depth of H$\beta$ compared to that of H$\alpha$. If the second peak in H$\alpha$ of our observed flare ($t \ \sim \ 40$ min) was a non-white-light flare, the optical depth of H$\beta$ must have been relatively high.

The second is the contribution from postflare loops. \citet{Otsu et al. 2024} conducted the Sun-as-a-star analysis of an X-class flare on the Sun to investigate the dynamics of postflare loops in spatially integrated data. They observed an X1.6 flare with the Solar Dynamics Doppler Imager (SDDI, \cite{Ichimoto et al. 2017}) mounted on the Solar Magnetic Activity Research Telescope (SMART, \cite{Ueno et al. 2004}) at Hida Observatory, Kyoto University. The observed flare occurred near the west-side solar limb (see figure 1 of \cite{Otsu et al. 2024}). 

Figure 4(a-2) in \citet{Otsu et al. 2024} presents the light curve of the EW of H$\alpha$ from spatially integrated data. The figure shows two distinct peaks. The first peak of H$\alpha$ is at the same time as the peak of GOES X-ray light curve. The second peak of H$\alpha$ is delayed $\sim \ 13$ min from the first peak. These peaks correspond to the flare ribbons and the postflare loops, respectively. After that, the off-limb loops emerged and caused additional brightness. As a result of that, the EW of H$\alpha$ does not monotonically decrease as in GOES but stops at around $t \ \sim \ 150$ min in figure 4(a-2) in \citet{Otsu et al. 2024}. 

If we apply the results of solar observations in \citet{Otsu et al. 2024} to our event on the M-type star, we can confirm similar features to figure 4(a-2) in \citet{Otsu et al. 2024} from flux of H$\alpha$ in figure \ref{fig:Balmer_flux_and_ratio}(a) in this paper: the first peak ($t \ \sim \ 10$ min), and the second peak ($t \ \sim \ 40$ min). By comparing our data with the results of \citet{Otsu et al. 2024}, the first peak and the second peak in figure \ref{fig:Balmer_flux_and_ratio}(a) are derived from flare ribbons and postflare loops, respectively. Furthermore, figure \ref{fig:Balmer_flux_and_ratio}(a) also shows the gradual increase after $t \ \sim \ 100$ min. In the case of \citet{Otsu et al. 2024}, the decrease in the EW of H$\alpha$ stops due to the off-limb loops. In our event, if the loops are brighter or larger than the loops reported in \citet{Otsu et al. 2024}, it is even possible for the decay not only to stop but to turn into a brightening instead. After that, the flux of H$\alpha$ decreases again. This may be because the density of plasma decreased and/or loops moved to the far side of the disk. However, observation of \citet{Otsu et al. 2024} ended and did not detect this kind of decrease. For a more detailed interpretation, we need longer-term and more diverse Sun-as-a-star observations. In addition, the possibility of the postflare loop can be also supported by the spot properties on the time of the flare. According to \citet{Ikuta et al. 2023}, one of spots is derived to be near the east-side stellar limb at the rotational phase of the flare (= 0.984) for both cases. Again, it is noted that this interpretation is based on the assumption that the results from solar flares can be applied to stellar flares on M-type stars.


Another aspect supporting the detection of postflare loops is the flux ratio of H$\beta$ to H$\alpha$. Figure \ref{fig:Balmer_flux_and_ratio}(b) shows the time variation of H$\beta$/H$\alpha$ ratio.
At the flare peak, the H$\beta$/H$\alpha$ ratio increased sharply. When the flare occurred, the optical depth of the chromospheric plasma increased. H$\beta$ is emitted from optically thin chromospheric plasma relative to H$\alpha$. H$\beta$ receives more contribution from plasma heated deeper in the chromosphere, making it more sensitive to flare heating than the H$\alpha$ line. This sharp increase in the flux ratio can be explained by the flare heating in the deep chromosphere. On the other hand, there is no significant indication of the second peak in the ratio of H$\beta$/H$\alpha$ in figure \ref{fig:Balmer_flux_and_ratio}(b). This suggests that the radiative mechanisms responsible for the Balmer lines at the flare peak and around the second peak are different. The absence of substantial heating in the deeper layers indicates that the emission may not originate from radiation from flare ribbons, but rather from post-flare loops. If the second peak of H$\alpha$ derives from non-white-light flare, there might be a significant signal of flux ratio of H$\beta$/H$\alpha$ at $t \ \sim \ 40$ min.

For these reasons, we concluded that the observed flare was likely accompanied by postflare loops and occurred near the west-side stellar limb. It would be preferable to have additional information to further constrain the flare location, such as Doppler velocity measurements. Since the projected rotational velocity, $v \sin i$, of EV Lac is $3.5 \ \mathrm{km/s}$ (\cite{Reiners et al. 2018}), we were unable to detect any Doppler shift associated with the flare due to the low dispersion of VPH-blue ($\lambda \ / \ \Delta\lambda \ \sim 500$). For future studies, high-time-cadence and high-dispersion observations will be necessary to investigate the detailed flare processes. Also, in order to investigate flare loops on stellar flares quantitatively, it is necessary to calculate the radiation from flare loops, assuming some parameters such as the loop size. 

\section{Summary and future work}
In this study, we performed simultaneous photometric and spectroscopic observation of the active M-dwarf EV Lac with TESS and Seimei Telescope. As a result, we detected a white-light flare simultaneously on 2019 September 14, at BJD 2458741.03. 

We estimated the temperature of the flare component, finding that the temperature at the flare peak was $8122 \ \pm \ 273$ K, which is comparable to results in previous studies. Additionally, we estimated the temperature of all spectra of the flare component throughout the flare. We confirmed that the e-folding time of the temperature and optical continuum flux were 2.5 min and 30 min, respectively, indicating that the flare temperature decayed more rapidly than the optical continuum flux. For instance, $\sim \ 10$ min after the flare peak, the flare temperature had decreased by about half. 

Using this time variation in the flare temperature ($T_{\mathrm{BB}}(t)$) and the flare area ($A_{\mathrm{flare}}(t)$) obtained from spectroscopic observations, we calculated the radiative flare energy within the optical wavelength range, obtaining a value of $4.4 \ \times \ 10^{32}$ erg. We also estimated the flare energy under the assumption that the flare component behaved as a fixed single-temperature blackbody radiator throughout the flare and the flare area ($A_{\mathrm{flare}}(t)$) was time-dependent in order to explain the TESS data. The flare energy for $T_{\mathrm{BB}} = 8122$ K was $8.9 \ \times \ 10^{32}$ erg, while for $T_{\mathrm{BB}} = 10000$ K, it was $1.2 \ \times \ 10^{33}$ erg. Since the ratio of flare energy within TESS-band to bolometric flare energy exhibits a temperature dependence (e.g., \cite{Schmitt et al. 2019}), if the time variation of flare temperature is ignored, the energy can be overestimated by a factor of 2 to 3.

We can confirm three features from the light curve of the Balmer lines: a delayed increase relative to the optical continuum flux ($t \ \sim \ 10$ min), a secondary increase ($t \ \sim \ 40$ min), and a gradual increase after the white-light flare ended (from $t \ \sim \ 100$ min). 
We compared our light curve with observational data from a solar flare that occurred near the west-side solar limb and was accompanied by postflare loops \citep{Otsu et al. 2024}. The light curve of the solar flare exhibited the same features as our light curve, suggesting that we may have detected postflare loops in the stellar flare. This is also supported by the fact that one of spots is derived to be near the stellar limb on the time of the flare from the TESS light curve \citep{Ikuta et al. 2023}.

For future work, we need to investigate the continuum emission during white-light flares in detail. We should compare observational data with the results of radiative transfer or model simulation. \citet{Kowalski et al. 2024} used the RADYN code to calculate a grid of 1D radiative-hydrodynamic stellar flare models that are driven by various electron-beam heating parameters and investigated how white-light continuum depend on the input electron-beam parameters (e.g., see figures 3 \& 5 in \cite{Kowalski et al. 2024}). \citet{Heinzel 2024} calculated one-dimensional radiative transfer of the hydrogen recombination continuum in stellar flares on M-and G-type stars. Figure 1 in \citet{Heinzel 2024} shows the simulated spectra from NUV to NIR. The colored spectra in figure 1 of his paper correspond to $T \ = \ 10000$ K, while their electron density differs from each other. To estimate the flare temperature, we assumed blackbody radiation within $4300$ $\leqq \ \lambda \ \leqq \ 7500$ \AA\, in subsection 3.2. Within this range, it is difficult to distinguish whether changes in the spectral slope are due to variations in flare temperature or column density. To separate the effect of temperature and column density, we should observe the discontinuities at $\lambda \ \sim \ 3600$ \AA\, and $\lambda \ \sim \ 8200$ \AA, known as the Balmer jump and Paschen jump, respectively. 

Moreover, our results will contribute to research about close-in habitable exoplanets. The time evolution of flare loops will help to investigate how EUV radiation from flare loops affect the atmospheric escape of exoplanets. To gain more insight into the flare loops, we will conduct time-resolved observations on stellar and solar flares and collect more observational cases. To investigate the impact of flares on photodissociation processes of molecules, we will also examine the continuum radiation from optical to NUV/FUV.

\begin{ack}
We thank Petr Heinzel, Shun Inoue, and Ayumi Asai, for their comments and discussions. This research was supported by the Japan Society for the Promotion of Science (JSPS) KAKENHI grant. 
JP20H05643(H.M.), JP21H01131(D.N., H.M., and K.S.), JP21J00316(K.N.), JP22K03685(H.M.), JP23K17694(K.S.), JP24H00248 (D.N., K.N., H.M., K.I., and S.H.), JP24K00680 (D.N., K.N., H.M., S.H., and K.S.), JP24K00685 (H.M.), and JP24K17082 (K.I.). This work was also supported by JST, the establishment of university
fellowships towards the creation of science technology
innovation, Grant Number JPMJFS2123 (T.O.).
The spectroscopic data used in this paper were obtained through the program 19B-NCN05 (PI: K.N.) with the 3.8 m Seimei telescope, which is located at Okayama Observatory of Kyoto University. This paper includes data collected with the TESS mission, obtained from the MAST data archive at the Space Telescope Science Institute (STScI). Funding for the TESS mission is provided by the NASA Explorer Program. STScI is operated bu the Association of Universities for Research in Astronomu, Inc., under NASA contact NAS 5-26555. Some of the data presented in this paper were obtained from MAST at the Space Telescope Science Institute. 

\end{ack}

\appendix 
\section{Airmass during Seimei observation}
\begin{figure}[h]
 \begin{center}
  \includegraphics[width=8cm]{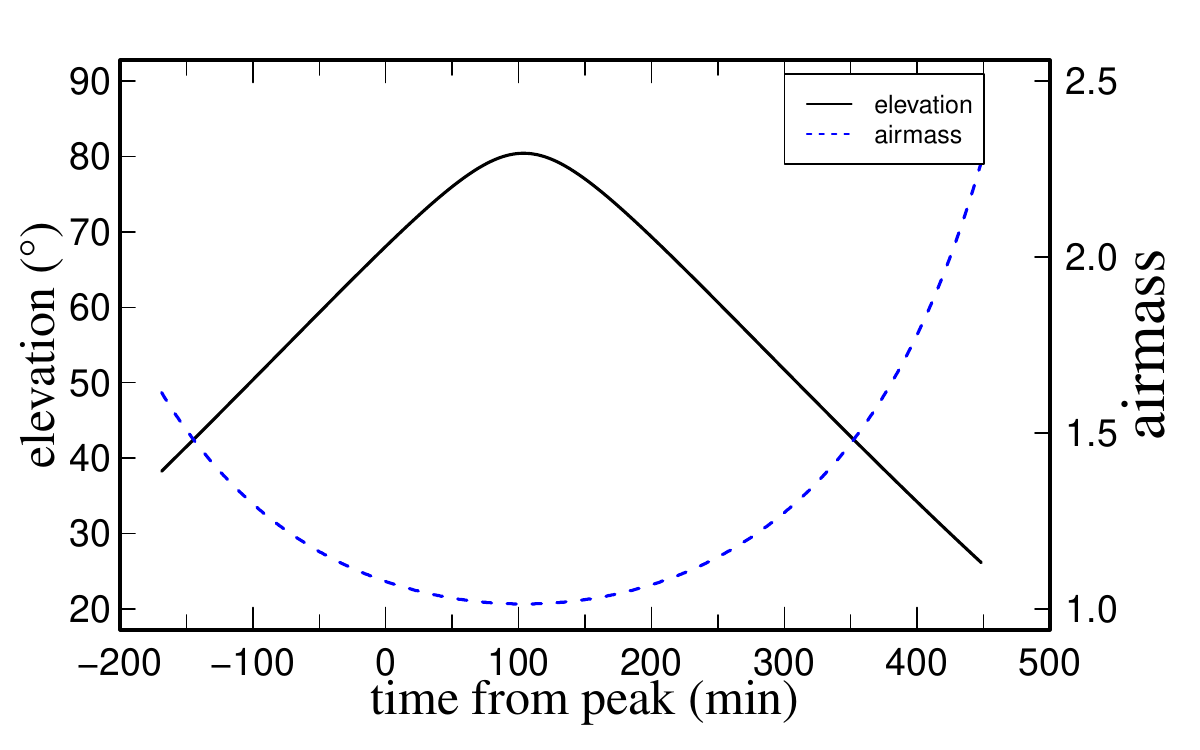} 
 \end{center}
\caption{ The value of the elevation of EV Lac and corresponding airmass during Seimei observation. The horizontal axis represents the time from photometric flare peak in minutes. The left-hand vertical axis shows the elevation of Seimei telescope in degree. The right-hand vertical axis shows the airmass. Black solid line and blue dashed line indicate the elevation and the airmass, respectively.
{Alt text: Line graph. The x axis represents the time relative to the photometric flare peak from -200 to 500 minutes. The left-hand y axis shows elevation of EV Lac from 20 to 90 degrees. The right-hand y axis shows airmass from 1.0 to 2.5.} 
}\label{fig:airmass}
\end{figure}

In this section, we show the elevation of EV Lac and airmass during whole Seimei observation. Figure \ref{fig:airmass} shows the time variation of the elevation of EV Lac and corresponding airmass during Seimei observation.

\section{Detrending long-term trend of the TESS light curve}
\begin{figure*}[h]
 \begin{center}
  \includegraphics[width=16cm]{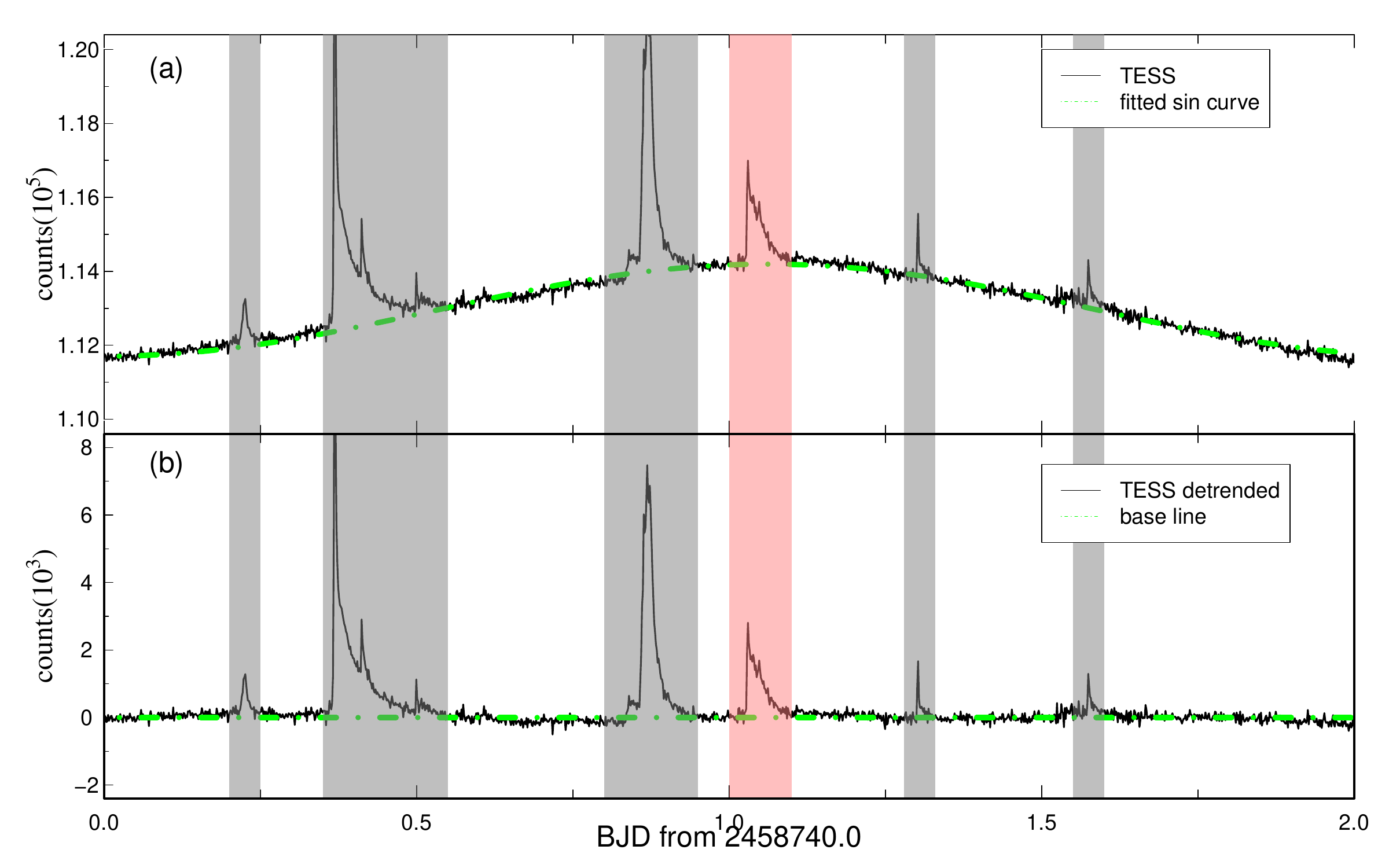} 
 \end{center}
\caption{ TESS light curve. The horizontal axis represents the time BJD from 2458740.0. The left-hand vertical axes show raw counts of TESS. Gray and red shaded areas are not used for sine fitting. Red shaded area include the flare we discussed in this paper. (a)Solid line indicates the TESS photometric light curve, and green dash-dotted line indicates fitted sine curve. (b)Solid line indicates the detrended TESS light curve, and green dash-dotted line indicates the base line.
{Alt text: Two line graphs. The x axis represents the time in BJD - 2458740.0 from 0.0 to 2.0. In the upper panel, the y axis shows raw counts of TESS from 1.10 to 1.20 times ten to the fifth. In the lower panel, the y axis shows raw counts of TESS from -2 to 8 times ten to the fourth.} 
}\label{fig:detrend}
\end{figure*}

To remove the long-term trend of the TESS light curve, we fitted single sine curve to the light curve. Figure \ref{fig:detrend}(a) shows the light curve of TESS and fitted sine curve, and figure \ref{fig:detrend}(b) shows the detrended TESS light curve.

\end{document}